# Empirical model of the Gibbs free energy for saline solutions of arbitrary concentration: Application for $H_2O$–NaCl solutions at 423.15 K-573.15 K and pressures from saturation up to 5 kbar

Mikhail V. Ivanov[*] and Sergey A. Bushmin

*Institute of Precambrian Geology and Geochronology, Russian Academy of Sciences. Nab. Makarova 2, Sankt-Petersburg, 199034, Russia*

**Abstract.**

An empirical model of the concentration dependence of the excess Gibbs free energy $G^{ex}$ for saline solutions is proposed. Our simple analytical form of $G^{ex}$ allows obtaining equations of state of saline solutions equally precise in the whole range of the salt concentrations, from dilute solutions up to the limit of solubility. Our equation for $G^{ex}$ includes one term responsible for concentration dependence of $G^{ex}$ at low salt concentrations and two terms of Margules type dependent on powers of mole fractions of the components. These terms contain four parameters dependent on temperature and pressure. As an example of application of the proposed model we took the system $H_2O$–NaCl. For fixed $T$ and $P$, our four-parameter form of $G^{ex}$ allows precisely reproduce experimental data for NaCl water solutions, including both regions of low and high concentrations. An introduction of temperature dependence of the parameters allowed us to build a precise analytic fit of experimental data on osmotic coefficients of NaCl solutions for saturation pressures of NaCl at temperatures 423.15 K-573.15 K. On the basis of *PVTx* data by Driesner (2007) the model was extended towards pressures up to 5 kbar. Our form of $G^{ex}$ allows obtaining the molar volumes, activities of components and other thermodynamic properties of NaCl–$H_2O$ solutions in the range of temperatures 423.15 K-573.15 K and pressures from saturation up to 5 kbar in the whole range of concentrations possible for $H_2O$–NaCl solutions (from zero up to approximately 10 molal at 573.15 K).

---

[*] Corresponding author. *E-Mail address:* m.v.ivanov@ipgg.ru

# 1. Introduction

The knowledge of the thermodynamic behavior of aqueous salt solutions is a key element for understanding the physical-chemical state of the Earth's crust fluids. Many of associated geological problems relate to hydrothermal systems with the ubiquitous NaCl and, in particular, are bound with the domain of temperatures 423.15 K-573.15 K at pressures up to 5 kbar. One of these problems concerns, for example, the ore-forming hydrothermal systems of epithermal, some part of porphyry, and the most part of orogenic gold deposits (*e.g.* Groves *et al*, 1998; Kerrich *et al*, 2000; Heinrich, 2005; Goldfarb and Groves, 2015). Aqueous NaCl solutions are a key subsystem of a vast number of more complex systems (*e.g.* Duan and Mao, 2006; Akinfiev et al, 2016). Especially important is the system $H_2O$–NaCl–$CO_2$, which plays a very significant role in a broad variety of geological processes and is a key system in studies of fluid inclusions in minerals. Theoretical studies of thermodynamic properties of $H_2O$–NaCl solutions, in some cases conjugated with the study of triple system $H_2O$–NaCl–$CO_2$, were done by Rogers and Pitzer (1982), Pitzer *et al* (1984), Lvov and Wood (1990), Archer (1992), Anderko and Pitzer (1993), Duan *et al* (1995), Duan and Sun (2003), Anovitz *et al* (2004), Evans and Powell (2006), Mao *et al* (2010, 2015), and Dubacq *et al* (2013). Many other publications on the subject were analyzed by Driesner (2007). The system $H_2O$–NaCl plays an exceptional role in researches of electrolytes, being nowadays the best studied electrolyte, both experimentally and theoretically. This makes $H_2O$–NaCl the most natural object for development and applications of new thermodynamic models.

Historically, a very important role in development of theoretical descriptions of thermodynamics of electrolytes, including $H_2O$–NaCl, played the model by Pitzer (Pitzer, 1972; Pitzer *et al*, 1984), which allowed obtaining relatively simple and precise descriptions for many solutions at low and intermediate concentrations. Nevertheless, this model has substantial restrictions in the case of highly concentrated solutions. On the other hand, models precise for high concentrations appeared to be not satisfactory for dilute solutions (Palaban and Pitzer, 1990). Extensions of the Pitzer's model to the whole range of concentrations (Pitzer and Simonson, 1986; Clegg and Pitzer, 1992; Clegg *et al*, 1992; Rard *et al*, 2010) were associated with a significant increase of its complexity and were not widely used for geological natural brines. Few exceptions are the works by Anderko and Pitzer (1993) and Jiang and Pitzer (1996). More recent models both for NaCl solutions and mixtures of several salts (Springer *et al*, 2012, 2015; Dubacq *et al*, 2013; Hingerl *et al*, 2014) are also rather complicated and difficult for reproduction.

The direct purpose of the present work is development of a simply formulated empirical numeric thermodynamic model, which should describe salt solutions in a uniform way and equally precisely, independent on their concentrations from dilute up to saturated solutions. Bearing in mind geological applications for fluids with the ubiquitous NaCl at crustal temperatures and pressures and the fact, that any empirical model should be based on a sound set of experimental data, we have chosen the system $H_2O$–NaCl at temperature 423.15 K-573.15 K and pressures to 5 kbar as a basis for development and testing our model. A large number of experimental data on thermodynamic behavior of $H_2O$–NaCl solutions can be grouped into two sets, convenient for theoretical analysis: 1. Activity of water and osmotic coefficient at saturation pressure of pure water at various temperatures and concentrations of NaCl; 2. *PVTx* properties of saline solutions at pressures equal and higher than the saturation pressure. For experimental data on $H_2O$–NaCl in *TP* range of our interest there are precise fits of both the osmotic coefficient (Liu and Lindsay, 1972) and *PVTx* data (Driesner, 2007). This is an important circumstance, allowing us to separate the development of the model from the analysis of raw experimental data. We used the first set of experimental data for development of our analytical form of the excess Gibbs free energy in the part, concerning its dependence of the salt concentration. On the other hand, fitting our model on *PVTx* data allowed us to check a predictive potential of our approach and obtain a practical thermodynamic model of NaCl water solutions for pressures up to 5 kbar.

In this paper, we concentrate on the development and testing the model for $H_2O$–NaCl. In the following, the existence of a proven model will facilitate its applications to the other water-salt systems of geological value, like solutions of $CaCl_2$, KCl, and $MgCl_2$. Sets of experimental data for these systems are more restricted comparing to $H_2O$–NaCl. Therefore, these systems require the proved theoretical models for obtaining their thermodynamic description.

**2. Method**

The most convenient representation for the set of experimental data on $H_2O$–NaCl solutions at saturation pressures is their expression through the osmotic coefficient $\varphi$

$$\varphi = -(1000 / vmM_1) \ln a_1 \quad (1)$$

Here $M_1 = 18.01534$ g is molar mass of water, $m$ is molality of the $H_2O$–NaCl solution, and $v = 2$ is the number of ions into which NaCl molecules dissociate in the solution. Activity of water $a_1$ looks like

$$a_1 = \exp[(\mu_1 - \mu_{0,1})/RT] \tag{2}$$

where $\mu_1$ is the chemical potential of water in the solution, $\mu_{0,1}$ is the chemical potential of pure water, $R = 8.3144598 \text{ J/mol/K}$ is the gas constant, and $T$ is the temperature (here and everywhere in formulas in Kelvins). The chemical potential is defined by the well-known formula

$$\mu_i = \left(\frac{\partial \hat{G}}{\partial N_i}\right)_{T,P,N_{j\neq i}} \tag{3}$$

where $\hat{G}$ is the full Gibbs free energy of the solution, $N_i$ is the number of particles of kind $i$ in the solution, and $P$ is the pressure. For convenience we will use, in the following, the Gibbs free energy per one mole of solution $G$[J/mol] instead of $\hat{G}$

$$G = \hat{G}/n \tag{4}$$

where $n$ is the total number of moles in the solution. The Gibbs free energy $G$ is a sum of the Gibbs free energies of pure solvent and solute and the excess Gibbs free energy $G^{ex}$

$$G = G^{ex} + x_1 G_1^0 + x_2 G_2^0 \tag{5}$$

Here $x_1 = x_{H_2O}$ and $x_2 = x_{NaCl}$ are mole fractions of $H_2O$ and NaCl. $G_1^0 = G_{H_2O}^0$ and $G_1^0 = G_{NaCl,sol}^0$ are Gibbs free energies of one mole of pure water and one mole of pure dissolved NaCl at given temperature and pressure. The excess Gibbs free energy can be considered to consist of a part corresponding to an ideal solution $G^{id}$ and non-ideal part $G^{non-ideal}$, representing physical interaction of solvent and solute molecules

$$G^{ex} = G^{id} + G^{non-ideal} \tag{6}$$

The ideal part of $G^{ex}$ has the well-known form

$$G^{id} = RT(x_1 \ln x_1 + x_2 \ln x_2) \tag{7}$$

Our goal is development of an analytic form of the excess Gibbs free energy able to reproduce existing experimental data and, thus, having a potential for obtaining quantities, not known from experiments. As an initial point, we have chosen the form of $G^{non-ideal}$ developed by Aranovich and Newton (1996, 1998) and finally formulated by Aranovich *et al* (2010) for the ternary system $H_2O$–$CO_2$–NaCl at high temperatures and pressures. For the subsystem $H_2O$–NaCl their non-ideal excess Gibbs free energy consists of two terms

$$G^{non-ideal,A} = G_\alpha + G_2 \tag{8}$$

where

$$G_\alpha = RT[\alpha x_2 \ln(\alpha x_2) - x_1 \ln(1 + \alpha x_2) - (1+\alpha) x_2 \ln(1 + \alpha x_2) + (1+\alpha) x_2 \ln(1 + \alpha)] \tag{9}$$

- 4 -

and
$$G_2 = x_1 x_2 W_2 = g_2 W_2 \tag{10}$$

with parameters $\alpha$ and $W_2$ dependent on temperature and pressure $P$. Initial meaning of the parameter $\alpha$ is the degree of dissociation of NaCl molecules (Aranovich and Newton, 1996, 1998) independent on the concentration of the solution. It should be noted that the introduction of term $G_2$ and other terms in $G^{ex}$ in addition to $G_\alpha$ deprives the parameter $\alpha$ of its meaning of the degree of dissociation at all the NaCl concentrations, where these additional terms are not equal to zero. Due to this reason and taking into account that for our temperature range 423.15-573.15K it is reasonable to suppose the full dissociation of NaCl at its low concentrations, this parameter was fixed as equal to one, $\alpha = 1$, $\alpha + 1 = \nu$. The chemical potential of water corresponding to $G^{\text{non-ideal,A}}$ is $\mu_1 = \mu_{1,\alpha} + \mu_{1,2}$ with partial chemical potentials given by Eqs. (A2) and (A4). This form of chemical potential appears to be insufficient for reproduction of relatively complex experimental dependence $\varphi(x_2)$ known for temperatures below the critical point of water.

Existing experimental data on the osmotic coefficient at 423.15-573.15K and pressure of saturation can be precisely reproduced when introducing two additional terms into $G^{ex}$, *i.e.*:

$$G_a = \left\{ x_2^{1/2} \ln\left(1 + x_2^{1/2} / \varepsilon_a\right) - x_2 \ln\left(1 + 1/\varepsilon_a\right) \right\} W_a = g_a W_a \tag{11}$$

and

$$G_6 = x_1 x_2^2 W_6 = g_6 W_6 \tag{12}$$

with parameters $\varepsilon_a$, $W_a$, and $W_6$. The second term in Eq. (11) is added for satisfying the requirement to the excess mixing energy being equal to zero for pure components.

The term $G_6$, containing $x_2^2$, is necessary for precise description of thermodynamics of the H$_2$O–NaCl solution at high concentrations of NaCl. The absence of this term or its equivalents in models like (Pitzer *et al*, 1984) and (Archer, 1992) causes the loss of their precision at high $x_{\text{NaCl}}$. The analogous problem in a more severe form exists in thermodynamic modeling for solutions of salts, better solvable than NaCl, for example CaCl$_2$ (Pitzer and Oakes, 1994).

Thus, our final form of the excess Gibbs free energy, applied in the following sections to the H$_2$O–NaCl solutions, is

$$G^{ex} = G^{id} + G_\alpha + G_a + G_2 + G_6 \tag{13}$$

Formulas for chemical potentials corresponding to the terms in Eq. (13) are given in the Appendix. The total chemical potential of water contains also a term coming from $x_1 G_1^0 + x_2 G_2^0$ and looks like

$$\mu_1 = G_1^0 + \mu_1^{id} + \mu_{1,\alpha} + \mu_{1,a} + \mu_{1,2} + \mu_{1,6} \tag{14}$$

At $x_2 = 0$ all the terms except $G_1^0$ are equal to zero. Thus

$$\mu_{0,1} = G_1^0 \tag{15}$$

Analogously, for the NaCl mole fraction mean ionic activity coefficient $\gamma_\pm$ we have

$$\gamma_\pm = \frac{1}{x_2} \exp\left(\frac{\mu_2 - \mu_{0,2}}{\nu RT}\right) \tag{16}$$

with

$$\mu_2 = G_2^0 + \mu_2^{id} + \mu_{2,\alpha} + \mu_{2,a} + \mu_{2,2} + \mu_{2,6} \tag{17}$$

and

$$\mu_{0,2} = G_2^0 + \mu_{0,2,\alpha} + \mu_{0,2,a} + \mu_{0,2,2} \tag{18}$$

with terms given in the Appendix. Both $G_1^0$ and $G_2^0$ do not affect the values of $a_1$, $\varphi$, and $\gamma_\pm$.

The equations given above form our thermodynamic model for NaCl water solutions. The equations contain four parameters $W_2$, $W_6$, $W_a$, and $\varepsilon_a$, which do not depend on the composition of the solution, but may depend on pressure and temperature. In the next section we consider our theoretical dependencies $\varphi(x_2)$ in comparison with the experimental data at fixed *TP* values.

**Table 1.** Parameters of our model obtained for steam saturation pressures at several temperatures.

| $T$ | $W_2$ | $W_6$ | $W_a$ | $\varepsilon_a$ |
|---|---|---|---|---|
| 423.15K | -2.91977096E+04 | -2.10731376E+03 | 8.34714300E+02 | 8.61959236E-02 |
| 448.15K | -2.08653196E+04 | 6.42458733E+03 | 1.02258911E+03 | 8.94195174E-02 |
| 473.15K | -1.41896446E+04 | 1.22444197E+04 | 1.23278655E+03 | 9.17406872E-02 |
| 498.15K | -8.83022688E+03 | 1.58095399E+04 | 1.48360483E+03 | 9.37999132E-02 |
| 523.15K | -7.10584288E+03 | 1.40032591E+04 | 1.74488231E+03 | 9.40423670E-02 |
| 548.15K | -8.45619296E+03 | 7.59199378E+03 | 2.05645257E+03 | 9.35785190E-02 |
| 573.15K | -1.51370430E+04 | -6.55208598E+03 | 2.41238084E+03 | 9.13955017E-02 |

## 3. Concentration dependencies of the osmotic coefficient

In this section, we rely on experimental data on osmotic coefficients of NaCl solutions at saturation pressures. On this basis, we fit parameters of our model, and check compatibility of our model with the results of experiments.

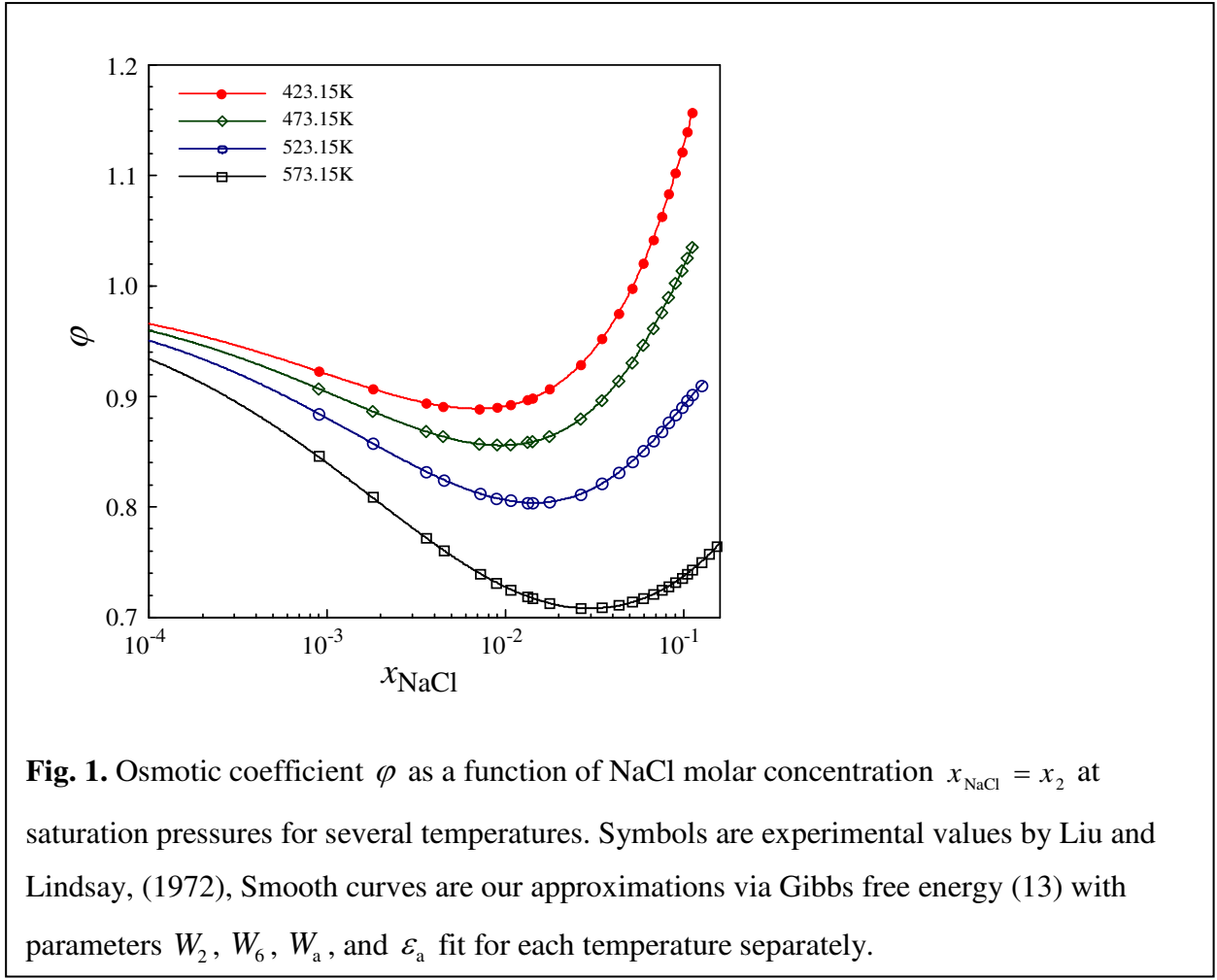

**Fig. 1.** Osmotic coefficient $\varphi$ as a function of NaCl molar concentration $x_{NaCl} = x_2$ at saturation pressures for several temperatures. Symbols are experimental values by Liu and Lindsay, (1972), Smooth curves are our approximations via Gibbs free energy (13) with parameters $W_2$, $W_6$, $W_a$, and $\varepsilon_a$ fit for each temperature separately.

Liu and Lindsay (1970, 1972) carried out an extensive analysis of their own and other experimental results on osmotic coefficients $\varphi$ for the system $H_2O$–NaCl within the temperature range 423.15 K-573.15 K at pressures of equilibrium between pure water and steam. For these *TP* conditions they provided a smooth approximation of $\varphi$ values, more useful than raw experimental data (Liu and Lindsay, 1972). Values of parameters $W_2$, $W_6$, $W_a$, and $\varepsilon_a$, obtained in our fitting on data by Liu and Lindsay (1972) for several temperature points are given in Table 1. The corresponding dependencies $\varphi(x_2)$ along with data by Liu and Lindsay (1972) are presented in Fig 1. We see a very precise reproduction of the experimental data with our curves originated from our representation of the Gibbs free energy (13) obtained by fitting only four numerical parameters.

A comparison of our results on the osmotic coefficient with the known works by Pitzer *et al* (1984) and Archer (1992) is provided in Fig 2. For the saturation pressure considered in this section we see, that for high NaCl concentrations (Figs. 2a,2b) our form of the Gibbs free energy is in a better agreement with the experimental data and have a right asymptotic behavior at high $m$ (or high $x_{NaCl}$). This precise reproduction of the experimental data is due to the presence of two power series terms $G_2$ and $G_6$ in the Gibbs free energy (13). These terms are similar to those of Margules decomposition of the Gibbs free energy. Terms $G_2$ and $G_6$ are neglectably small at low NaCl concentrations, but they provide precise representation of the thermodynamics of the solution at high concentrations of the salt. On the other hand, the difference between our (and experimental) results and Pitzer model at low molalities (Figs. 2c, 2d) is rather small at pressures of saturation.

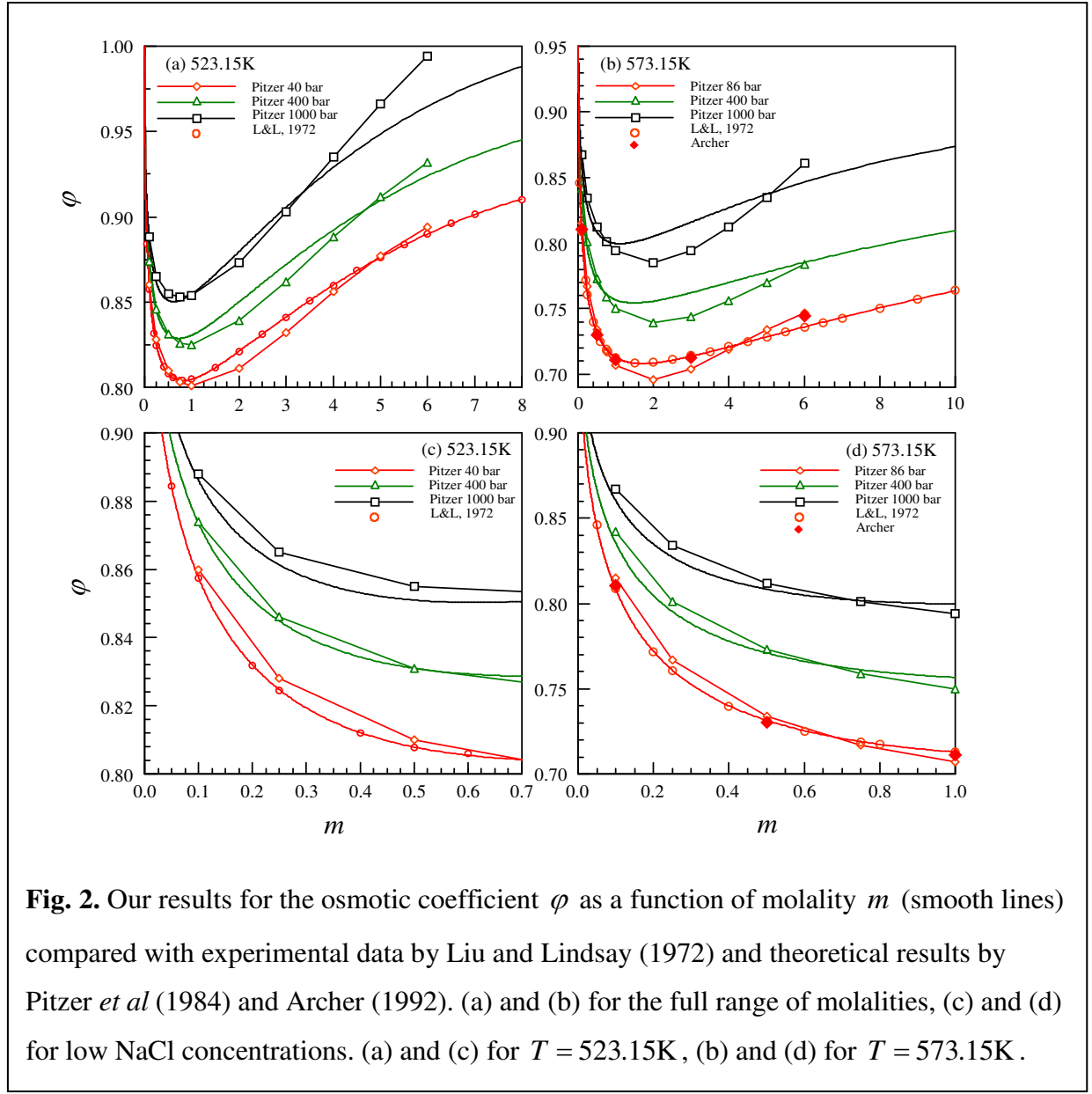

**Fig. 2.** Our results for the osmotic coefficient $\varphi$ as a function of molality $m$ (smooth lines) compared with experimental data by Liu and Lindsay (1972) and theoretical results by Pitzer *et al* (1984) and Archer (1992). (a) and (b) for the full range of molalities, (c) and (d) for low NaCl concentrations. (a) and (c) for $T = 523.15K$, (b) and (d) for $T = 573.15K$.

In our model, the behavior of the osmotic coefficient in dilute solutions is governed by the terms $G_\alpha$ and $G_a$. The value of the osmotic coefficient $\varphi \to 1$ at $x_2 \to 0$. This is provided by the form of $\mu_{1,\alpha}$ in the Eq. (A2). The behavior of the value $1-\varphi$ at small $x_2$ is predominantly regulated by the term $\mu_{1,a}$ (A3). Two leading terms in the power series decomposition of $1-\varphi$ at small $x_2$ look like

$$1-\varphi = \frac{W_a}{vRT}\left(\frac{1}{4}\frac{x_2^{1/2}}{\varepsilon_a^2} - \frac{1}{3}\frac{x_2}{\varepsilon_a^3} + \cdots\right) \tag{19}$$

This form of dependence is similar to Brønsted (1922) empirical equation (see also Guggenheim and Turgeon, 1955; Elizalde and Aparicio, 1995). At higher $x_2$ the behavior of the osmotic coefficient is determined by an interplay of all the terms in $G^{ex}$. At high $x_2$ values the term $G_6$ plays an important role in forming the shape of the curve $\varphi(x_2)$.

## 4. Temperature dependence of the osmotic coefficient at the vapor saturation pressure

The next step in the development of our model should be obtaining temperature and pressure dependencies of parameters $W_2$, $W_6$, $W_a$, and $\varepsilon_a$. Experimental data by Liu and Lindsay (1972) were obtained for pressures of equilibrium between liquid water and water steam. This pressure depends on the temperature. The latter means that parameters, obtained for the curves in Fig. 1, do not allow extracting their pure temperature dependencies at fixed pressures. Moreover, the range of pressures ($1-86\,\text{bar}$) covered by these experimental data is too narrow for obtaining pressure dependencies for many geological applications, requiring pressures up to several kilobars.

Our calculations show that temperature dependencies of parameters $W_2$, $W_6$, $W_a$, and $\varepsilon_a$ at corresponding saturation pressures can be precisely approximated by simple analytic formulas:

$$W_2(T) = W_{2T} = u_{20} + u_{22}(T-u_{21})^2 T^{3/2} \tag{20}$$

$$W_6(T) = W_{6T} = u_{60} + u_{62}(T-u_{61})^2 T^{3/2} \tag{21}$$

$$W_a(T) = W_{aT} = (T/u_{a0})^{7/2} \tag{22}$$

$$\varepsilon_a(T) = u_{\varepsilon 0} + u_{\varepsilon 2}(T-u_{\varepsilon 1})^2 T \tag{23}$$

Numerical parameters present in formulas (20)-(23) are given in Table 2 (column "Saturation pressure"). It should be noted, that these values are obtained in a fit, based on a large number of $\varphi(x_2, T)$ experimental data for various $x_2$ and temperatures. This fit is fully independent on the fixed $T$ fits, presented in the previous section. Therefore, Eqs. (20)-(23) do not precisely lead to values in Table 1. On the other hand, temperature dependencies of values of Table 1 allowed us to find the analytic form of Eqs. (20)-(23).

Our model values of $\varphi(x_2, T)$ reproduce experimental results by Liu and Lindsay (1972) with a precision not much different from that of $\varphi(x_2)$ obtained in the previous section. In the scale of Figure 1 both sets of curves coincide. In Figure 2 we show concentration dependencies of the osmotic coefficient $\varphi(x_2)$ following from fit of this section in comparison with experimental data and theoretical curves from the model by Pitzer *et al* (1984).

**Table 2.** Numerical parameters of the fit of temperature dependencies (20)-(23) of the Gibbs free energy at corresponding saturation pressures and same parameters for approximation (28)-(32) ($P = 0$).

|  | Saturation pressure | $P = 0$ |
|---|---|---|
| $u_{20}$ | -7.40465984E+03 | -7.23315191E+03 |
| $u_{21}$ | 5.26013556E+02 | 5.25693806E+02 |
| $u_{22}$ | -2.32223423E-04 | -2.29398977E-04 |
| $u_{60}$ | 1.51261293E+04 | 1.49169544E+04 |
| $u_{61}$ | 5.02648252E+02 | 5.00430451E+02 |
| $u_{62}$ | -3.04159146E-04 | -3.05638058E-04 |
| $u_{a0}$ | 6.19119130E+01 | 6.18519160E+01 |
| $u_{\varepsilon 0}$ | 9.44196901E-02 | 9.54781985E-02 |
| $u_{\varepsilon 1}$ | 5.22720408E+02 | 5.16170457E+02 |
| $u_{\varepsilon 2}$ | -2.05279225E-09 | -2.53953712E-09 |

## 5. Temperature and pressure dependencies of the Gibbs free energy

The next step in the development of our model should be obtaining pressure dependencies of parameters $W_2$, $W_6$, $W_a$, and $\varepsilon_a$. Experimental data, related to saturation pressures, are not sufficient for separation of temperature and pressure dependencies of thermodynamic functions. In addition, they cover a very narrow range of pressures. Thus, obtaining pressure dependence of the Gibbs free energy and values derived from it, requires involving additional experimental information.

The pressure dependence of the Gibbs free energy can be obtained on the basis of the existing data on the molar volume $V$ of NaCl solutions dependent on $T$, $P$, and $x_{NaCl}$ by means of the following relation

$$\left(\frac{\partial G}{\partial P}\right)_{T,x_i} = V \tag{24}$$

The analysis of multiple of experimental data on the molar volume of NaCl solutions on temperature, pressure and NaCl content is not the goal of the present work. Such an analysis was performed recently by Driesner (2007) and Mao and Duan (2008). These authors also developed *PVTx* models for $H_2O$–NaCl fluids which give good approximations of the existing experimental data. Some modification of Driesner (2007) model was done by Mao *et al* (2015). The latter work contains also a list of other *PVTx* models for $H_2O$–NaCl fluids. For our purposes we used the formulas by Driesner (2007), because they provide correct data on molar volumes $V(x_2)$ of $H_2O$–NaCl solutions in a broad region of pressure values up to 5 kbar.

The formulas by Driesner (2007) cannot be directly incorporated into our representation of the Gibbs free energy and should be transformed into a form compatible with our representation of $G^{ex}$. From practical reasons we established this transformation on the basis of the excess molar volume of the solution $V^{ex}$

$$V^{ex} = \left(\frac{\partial G^{ex}}{\partial P}\right)_{T,x_i} = V - x_1 V_1^0 - x_2 V_2^0 \tag{25}$$

where $V_1^0$ and $V_2^0$ are molar volumes of pure water and pure dissolved NaCl at given temperature and pressure. The use of $V^{ex}$ allowed us to avoid an incorporation of any *PVT* model of pure water as well models of pure dissolved NaCl into our scheme for $G^{ex}$.

Original data by Driesner (2007) were obtained on the basis of the IAPS-84 equation of state of pure $H_2O$ (Kestin and Sengers, 1986). For the sake of compatibility, we used in Eq. (25) the same IAPS-84 equation of state of pure water for obtaining $V_1^0$. For the same purpose, equations for density of pure dissolved NaCl presented by Driesner (2007) were utilized for obtaining $V_2^0$. The pressure dependencies of $V^{ex}$ values obtained in such a way were the subject of the fitting.

For fixed mole fractions of NaCl and temperatures between 423.15 K and 573.15 K, the dependencies $V^{ex}(P)$ can be precisely approximated by a polynomial with respect to $P^{1/4}$

$$V^{ex}(P) = v_1 + v_2 P^{1/4} + v_3 P^{1/2} + v_4 P^{3/4} + v_5 P \tag{26}$$

For temperatures below 523.15 K even more simple approximation with $v_2 = v_4 = 0$ gives good results. Despite of that, only the form (26) was used in the following. The term in the Gibbs free energy corresponding to Eq. (26) is

$$G^{ex}(P) = \int_0^P V^{ex} dP = v_0 + v_1 P + \tfrac{4}{5} v_2 P^{5/4} + \tfrac{2}{3} v_3 P^{3/2} + \tfrac{4}{7} v_4 P^{7/4} + \tfrac{1}{2} v_5 P^2 \tag{27}$$

where $v_0$ is a part of $G^{ex}$ independent on pressure and (pure formally) corresponds to $P = 0$. This form (26) of approximate dependence on $P$ was applied to each of coefficients $W_2$, $W_6$, and $W_a$ separately. Concerning parameter $\varepsilon_a$ we assumed its independence on pressure. It follows from this, that concentration dependencies of $V^{ex}$ are analogous to those given by Eqs. (10)-(12). Thus, our approximation for $V^{ex}$ looks like

$$V^{ex} = g_a V_a^{ex} + g_2 V_2^{ex} + g_6 V_6^{ex} \tag{28}$$

$$V_i^{ex} = v_{i1} + v_{i2} P^{1/4} + v_{i3} P^{1/2} + v_{i4} P^{3/4} + v_{i5} P, \quad i = 2, 6, a. \tag{29}$$

Our final form of the excess Gibbs free energy for aqueous NaCl solutions is given by Eq. (13) with coefficients $W_i$:

$$W_i(T, P) = W_{iT} + W_{iP}, \quad i = 2, 6, a. \tag{30}$$

$$W_{iP} = v_{i0} + v_{i1} P + \tfrac{4}{5} v_{i2} P^{5/4} + \tfrac{2}{3} v_{i3} P^{3/2} + \tfrac{4}{7} v_{i4} P^{7/4} + \tfrac{1}{2} v_{i5} P^2 \tag{31}$$

where $W_{iT}$ are given by Eqs. (20)-(23).

Fitting $V^{ex}(P, x_2)$ at a number of fixed temperature values shows, that temperature dependence of all the parameters $v_{i1}...v_{i5}$ can be approximated in the form

$$v_{ij} = v_{ij1} + v_{ij2} \exp(v_{ij3} T), \quad i = 2, 6, a, \quad j = 1, 2, 3, 4, 5. \tag{32}$$

Values of $v_{ijk}$ were obtained by fitting a large number (more than 2000) of $V^{ex}$ values, obtained from formulas by Driesner (2007) at temperatures from $T = 423.15$K to $T = 573.15$K, pressures from $P = 0.1$kbar to $P = 5$kbar, and molalities from $m = 0.1$ to maximal possible at corresponding temperatures. Values of $v_{i0}$ do not appear in an explicit form in our approach for $V^{ex}$. Instead of introducing $v_{i0}$ into our computational scheme, we had re-fitted values $u_{ij}$ together with $v_{ijk}$ on a full set of data including both $V^{ex}(P, T, x_2)$ by Driesner (2007) and $\varphi(P, T, x_2)$ by Liu and Lindsay (1972). New values of $u_{ij}$ correspond to a formal pressure value $P = 0$. They are given in the right column of Table 2. Values of $v_{ijk}$ are given in Table 3. Units for values in Table 3 correspond to pressures in kilobars and $V^{ex}$ in cm$^3$/mol. For obtaining values of $W_{iP}$ in J/mol, their values from Eq. (31) with parameters from Table 3 should be multiplied by $10^2$.

**Table 3.** Parameters in Eqs. (32).

| | | | | | |
|---|---|---|---|---|---|
| $v_{211}$ | -1.05582866E+01 | $v_{212}$ | -3.15956089E+00 | $v_{213}$ | 8.73783182E-03 |
| $v_{221}$ | 8.29636849E+01 | $v_{222}$ | 1.61028006E+01 | $v_{223}$ | 8.56594936E-03 |
| $v_{231}$ | -6.58295556E+01 | $v_{232}$ | -2.91983780E+01 | $v_{233}$ | 8.38975538E-03 |
| $v_{241}$ | 2.81504859E+01 | $v_{242}$ | 2.08560152E+01 | $v_{243}$ | 8.31964580E-03 |
| $v_{251}$ | -7.20726810E+00 | $v_{252}$ | -5.04240049E+00 | $v_{253}$ | 8.33975892E-03 |
| $v_{611}$ | 8.14134328E+01 | $v_{612}$ | 1.35150164E+01 | $v_{613}$ | 8.12465369E-03 |
| $v_{621}$ | -1.48502455E+02 | $v_{622}$ | -5.40834994E+01 | $v_{623}$ | 8.28136869E-03 |
| $v_{631}$ | 1.05602397E+02 | $v_{632}$ | 8.75390026E+01 | $v_{633}$ | 8.29977211E-03 |
| $v_{641}$ | 2.57086544E+00 | $v_{642}$ | -6.03730864E+01 | $v_{643}$ | 8.31771812E-03 |
| $v_{651}$ | -1.32849786E+01 | $v_{652}$ | 1.48601379E+01 | $v_{653}$ | 8.34469399E-03 |
| $v_{a11}$ | 3.05629897E+00 | $v_{a12}$ | 9.02632003E-01 | $v_{a13}$ | 8.70560856E-03 |
| $v_{a21}$ | -2.59933704E+00 | $v_{a22}$ | -3.42526132E+00 | $v_{a23}$ | 9.25811805E-03 |
| $v_{a31}$ | -7.29935646E+00 | $v_{a32}$ | 5.48344102E+00 | $v_{a33}$ | 9.32421155E-03 |
| $v_{a41}$ | 8.39465275E+00 | $v_{a42}$ | -3.84185589E+00 | $v_{a43}$ | 9.27519927E-03 |
| $v_{a51}$ | -2.18109375E+00 | $v_{a52}$ | 9.70771091E-01 | $v_{a53}$ | 9.20372469E-03 |

In a graphical form, our approximation for $V^{ex}$ is compared with the data by Driesner (2007) in Fig. 3. In Fig. 3a we present dependencies of $V^{ex}$ on the pressure for a set of various temperatures and molalities. The differences between these two approximations do not exceed deviations between initial experimental results and the approximation by Driesner (2007). It should be noted that Eqs. (29)-(32) do not depend on $x_2$. Thus, all the dependencies of $V^{ex}$ on $x_2$ given by Eq. (28) follow from Eqs. (10)-(12), which were obtained for approximation of experimental dependencies $\varphi(x_2)$. In Fig. 3b we present concentration dependencies of $V^{ex}$ for several pressure values at $T = 523.15K$. We see here a precise reproduction of dependencies $V^{ex}(x_2)$ by a sum of three terms analogous to Eqs. (10)-(12). This fact is an important argument in behalf of our representation of the Gibbs free energy of mixing. This fact justifies as well a possibility to apply our formulas for osmotic coefficient and activities for all the pressures up to 5 kbar.

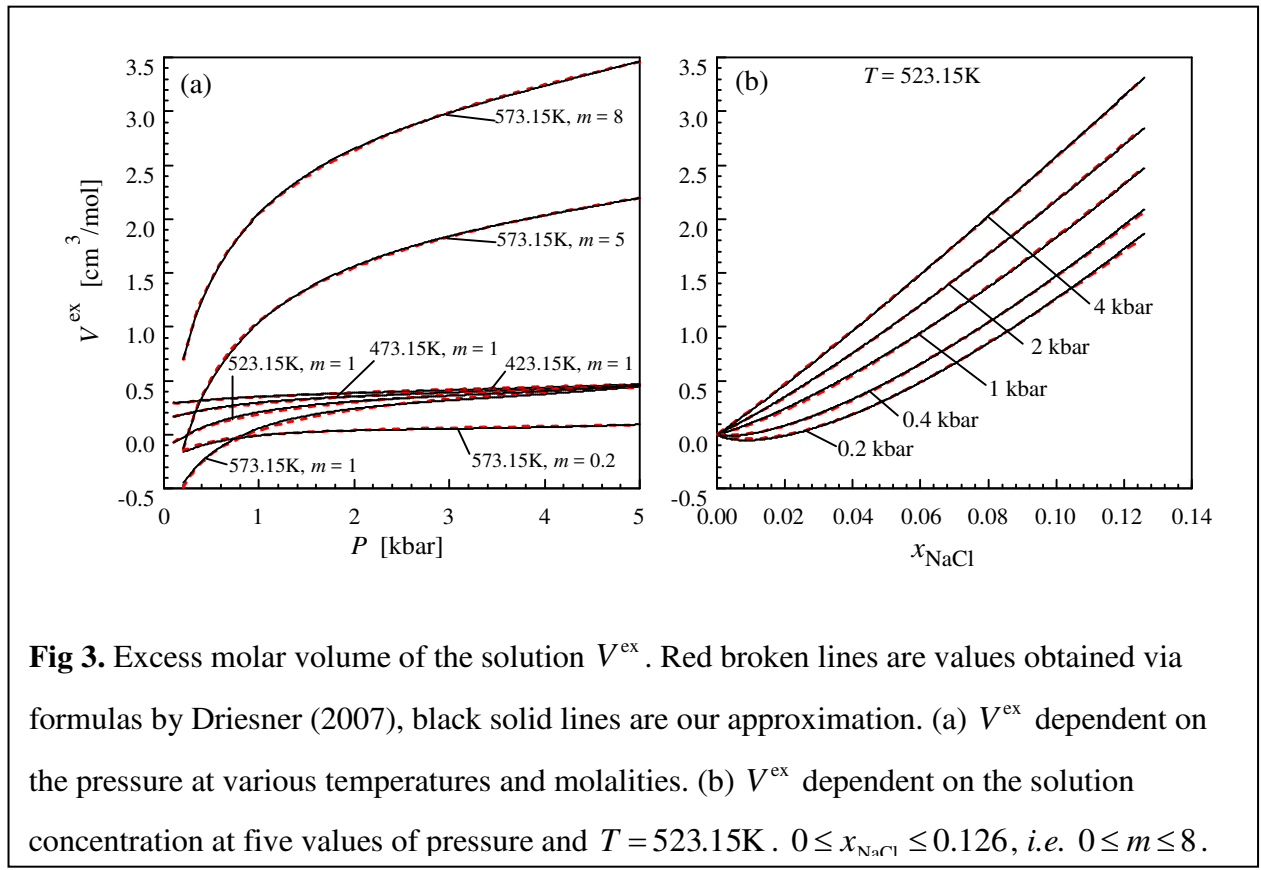

**Fig 3.** Excess molar volume of the solution $V^{ex}$. Red broken lines are values obtained via formulas by Driesner (2007), black solid lines are our approximation. (a) $V^{ex}$ dependent on the pressure at various temperatures and molalities. (b) $V^{ex}$ dependent on the solution concentration at five values of pressure and $T = 523.15\text{K}$. $0 \leq x_{NaCl} \leq 0.126$, *i.e.* $0 \leq m \leq 8$.

## 6. Discussion

Figs. 4-8 give a graphical representation of our basic results related to activities of components of H$_2$O-NaCl solutions. Dependencies of the osmotic coefficient on mole fraction of NaCl obtained via approximation presented above are presented in Fig. 4 for temperatures from 423.15 K to 573.15 K and pressures from the saturation pressure up to 5 kbar. For the saturation pressure, the osmotic coefficients resulting from the approximation, obtained in the previous section, are very close (with the average error less than $3 \times 10^{-4}$) to the experimental values by Liu and Lindsay (1972).

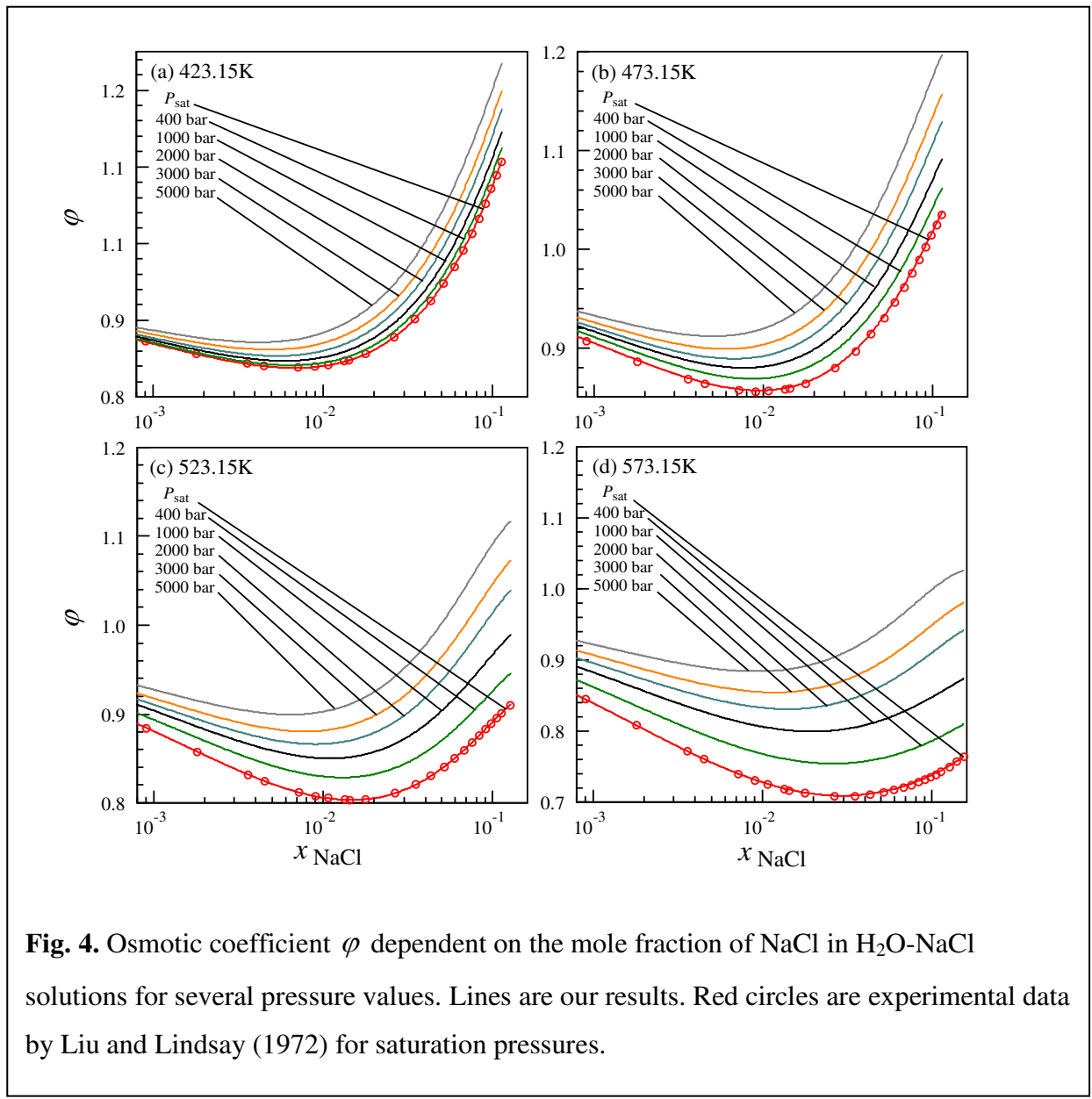

**Fig. 4.** Osmotic coefficient $\varphi$ dependent on the mole fraction of NaCl in $H_2O$-NaCl solutions for several pressure values. Lines are our results. Red circles are experimental data by Liu and Lindsay (1972) for saturation pressures.

Our values of the osmotic coefficient are compared with the results by Pitzer *et al* (1984) and Archer (1992) in Fig. 2. Discussion for saturation pressures is provided in Sec. 3. The discrepancies noted at high concentrations of NaCl are retained also for higher pressures. Values of $\varphi$, obtained in our calculations, increase with the pressure slightly slower than these by Pitzer *et al* (1984). The discrepancies in the osmotic coefficient values between theoretical and experimental data at high concentrations of NaCl are specific not only for early thermodynamic models by Pitzer *et al* (1984) and Archer (1992) but also most recent and well developed models. For example, osmotic coefficients by Sun and Dubessy (2012) increase with growing molality faster, than $\varphi$ calculated by Pitzer *et al* (1984) (Fig. 2). Thus, both these models show the same type of discrepance with experimental data by Liu and Lindsay (1972), more pronounced in the model by Sun and Dubessy, (2012). In addition, the range of prediction on pressures and NaCl concentrations in this model does not cover the whole possible range of molalities. The same relates to REUNIQUAC model by Hingerl *et al* (2014). It is restricted in values of molality of NaCl with $m \leq 5$, whereas our approach allows to cover the whole possible for $H_2O$–NaCl range of molalities $0 \leq m \leq 10$.

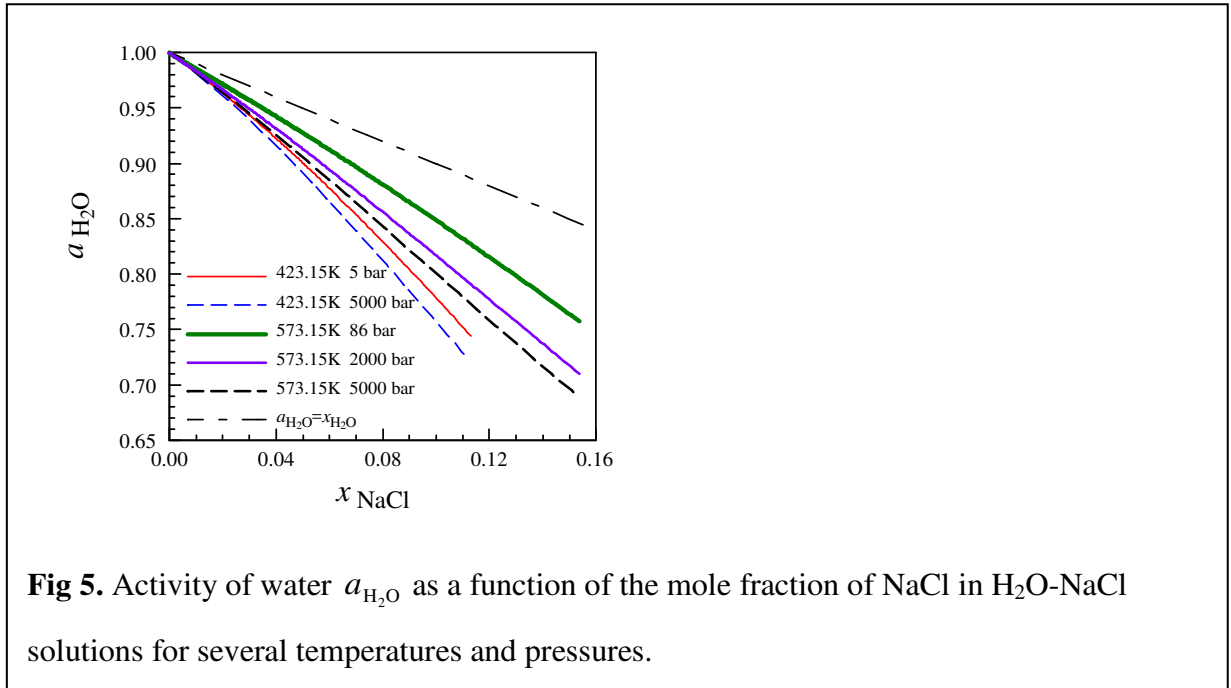

**Fig 5.** Activity of water $a_{H_2O}$ as a function of the mole fraction of NaCl in $H_2O$-NaCl solutions for several temperatures and pressures.

In Fig. 5 we present an example of activities of water $a_{H_2O}$ dependent on the mole fraction of NaCl. These dependencies are given for two temperature values and various pressures at these temperatures. These dependencies show that the activity of water decreases with growing pressure. The difference between high- and low-pressure $a_{H_2O}$ becomes more distinct with increasing temperature. The same tendency of decreasing $a_{H_2O}$ with $T$ was found experimentally by Aranovich and Newton (1996) for some higher temperatures. At the same time, we see an increase of $a_{H_2O}$ with temperature. This behavior looks to be in agreement with the experimental observation $a_{H_2O} \approx x_{H_2O}$ at temperatures between 590°C and 669°C and pressure 2 kbar (Franz, 1982; Aranovich and Newton, 1996).

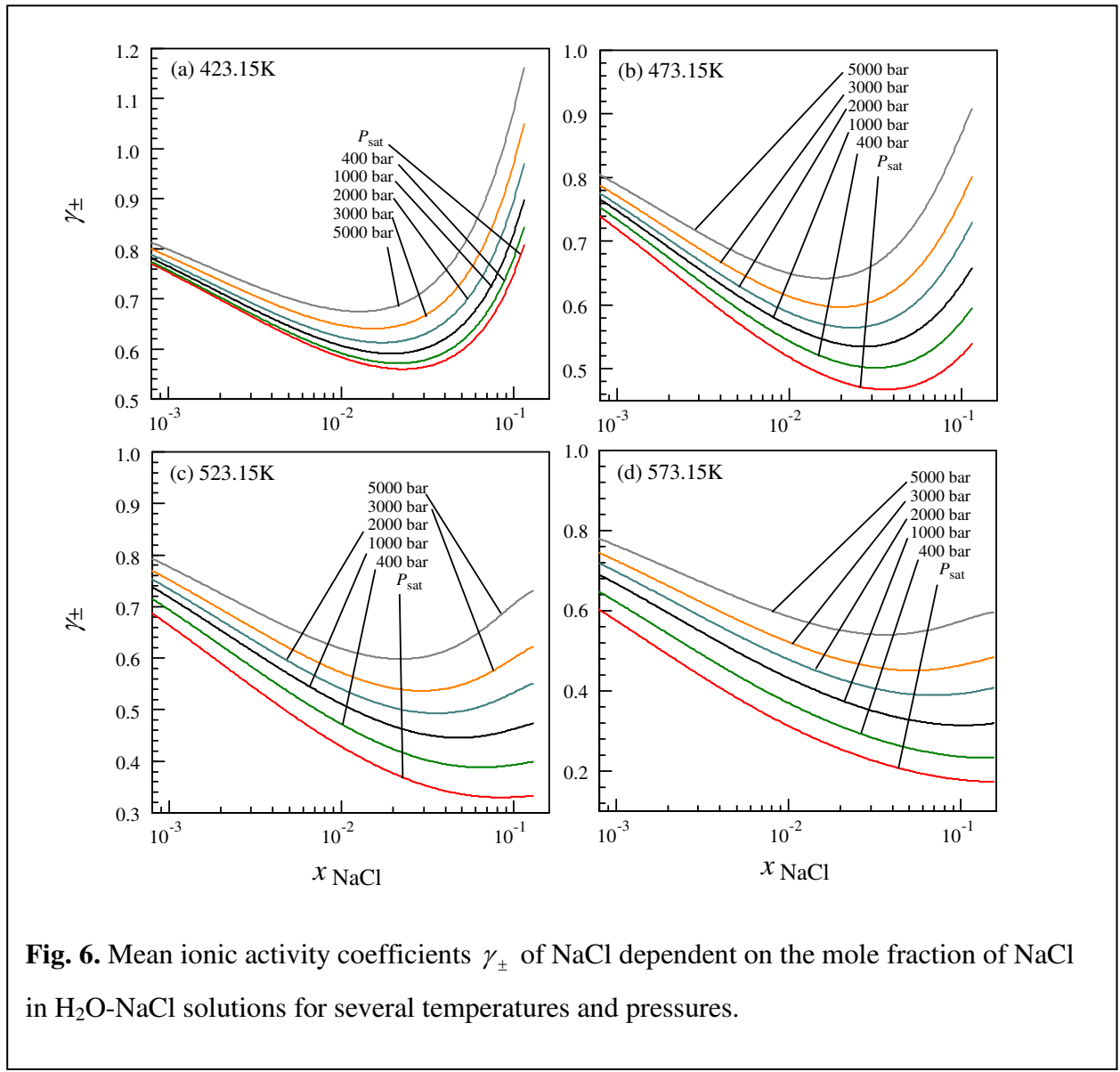

**Fig. 6.** Mean ionic activity coefficients $\gamma_\pm$ of NaCl dependent on the mole fraction of NaCl in H$_2$O-NaCl solutions for several temperatures and pressures.

In Fig. 6 we present concentration dependencies of the mean ionic activity coefficient $\gamma_\pm$ of NaCl for four temperatures from 423.15 K to 573.15 K and pressures from the saturation pressure to 5 kbar obtained with our thermodynamic model. In Fig. 7 we present a comparison of our results for (molal) mean ionic activity coefficient $\gamma_{\pm m}$ at saturation and elevated pressures with the models by Pitzer *et al* (1984) and Archer (1992). As well as for the osmotic coefficient, $\gamma_{\pm m}$ values in our model increase with $P$ slightly slower than in the model by Pitzer *et al* (1984). All the thermodynamic models, cited above are restricted in the range of molalities, where they can predict values of $\gamma_{\pm m}$. For models by Pitzer *et al* (1984), Archer (1992), and Sun and Dubessy (2012) this is $m \leq 6$. For Hingerl *et al* (2014) this is $m \leq 5$.

As it was mentioned in Sec. 2, our model is a kind of extension of models by Aranovich *et al* (Aranovich *et al*, 2010; Aranovich and Newton, 1996, 1998). Comparing to these models we added two terms to $G^{ex}$, *i.e.* term $G_a$, which reflects the long-range interaction of ions, and the second short-range term $G_6$. The two other most similar models of saline solutions published up to now are DH-ASF (Evans and Powell, 2006; Dubacq et al, 2013) and extended model of Pitzer (Pitzer and Simonson, 1986; Clegg and Pitzer, 1992; Clegg *et al*, 1992; Rard *et al*, 2010). For the short-range interaction of species the model DH-ASF contains terms, similar to $G_2 = x_1 x_2 W_2$, *i.e.* terms linear with respect to each separate mole fraction, but does not contain terms with higher degrees of $x_i$. Our model contains an additional short-range term $G_6 = x_1 x_2^2 W_6$, and, moreover, it enables introduction of additional terms of the form $x_1^{n_1} x_2^{n_2} W$ into $G^{ex}$ when this is necessary. This makes our model more flexible for precise reproduction of thermodynamic properties of solutions in a broad range of their concentrations. The extended model of Pitzer contains, in principle, the full set of terms with degrees of molar fractions of components. On the other hand, its, not very numerous, applications to natural brines are usually restricted by sets of fixed temperatures and pressures. Our model, presented in this paper covers a broad and important for geological applications range of pressures and temperatures. The second aspect, which distinguish our model from both the DH-ASF and extended model of Pitzer, relates to treating the long-range interactions of ions. Both these models consider it in the framework of the Debye-Hückel approximation, whereas we used for this purpose an empirical term $G_a$ with two parameters $\varepsilon_a$ and $W_a$. Our empirical approach proved to be at least not less precise for low NaCl concentrations than the Pitzer model, based on the Debye-Hückel theory (see Fig. 2). It is well known, that the Debye-Hückel theory is less precise for many electrolytes different from H$_2$O-NaCl. We expect that our simple model can be useful as an alternative approach for many salt solutions.

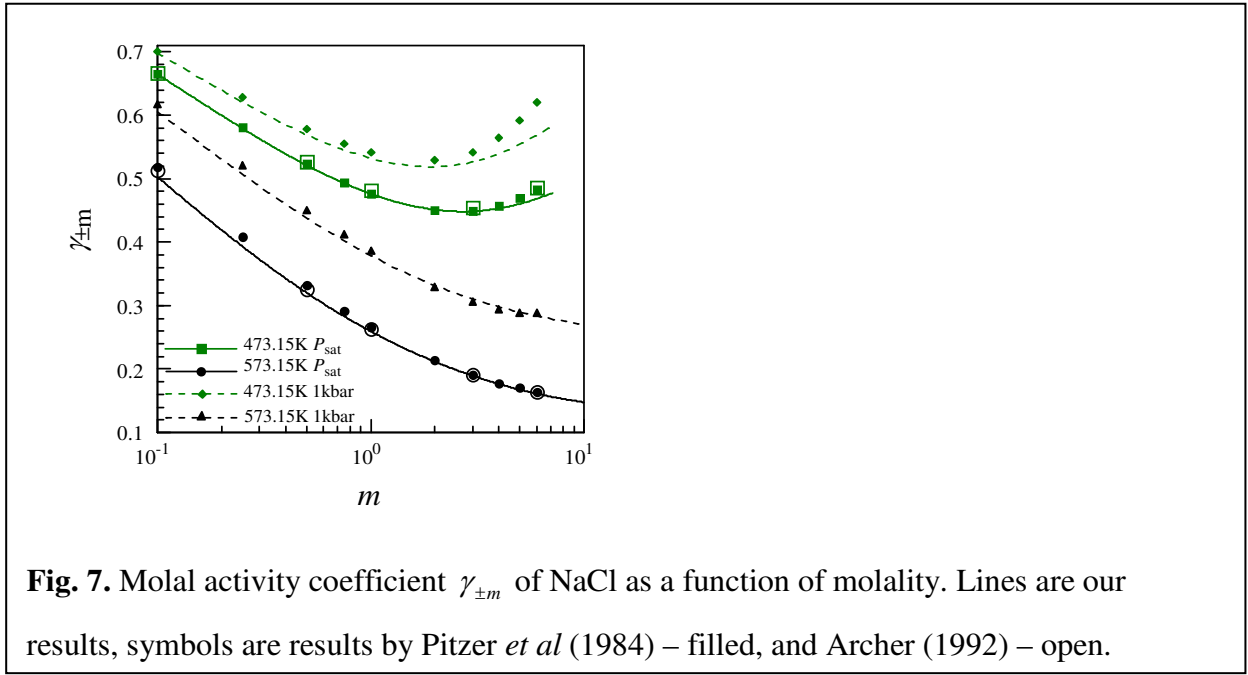

**Fig. 7.** Molal activity coefficient $\gamma_{\pm m}$ of NaCl as a function of molality. Lines are our results, symbols are results by Pitzer *et al* (1984) – filled, and Archer (1992) – open.

## 7. Conclusions

We have proposed an empirical model of the concentration dependence of the excess Gibbs free energy $G^{ex}$ for saline solutions. Our simple analytical form of $G^{ex}$ allows obtaining equations of state of saline solutions equally precise in the whole range of the salt concentrations, from dilute solutions up to the limit of solubility. The key feature of our approach is representation of the non-ideal part of the Gibbs free energy as a sum of terms of two kinds. One of these terms provides an adequate empirical representation of the Gibbs free energy for dilute solutions. The second kind of terms is responsible for thermodynamic behavior of the solution at intermediate and high concentrations of the salt. In the current paper, they are two terms of Margules type decomposition of the free energy on molar fractions powers. This set of terms contains four numerical parameters $W_2$, $W_6$, $W_a$, and $\varepsilon_a$. We have found that for fixed $T$ and $P$ this set of parameters is sufficient for precise thermodynamic description of NaCl–H$_2$O solutions at arbitrary concentrations accessible at subcritical temperatures. On the other hand, for other systems, the model allows evident extensions by including additional Margules-type terms.

We applied our approach to the system H$_2$O–NaCl at 423.15 K-573.15 K and pressures from saturation pressure up to 5 kbar. Introducing $TP$ dependencies of parameters $W_2$, $W_6$, $W_a$, and $\varepsilon_a$ we have build a precise numerical model of thermodynamic properties of the system. The model allow obtaining activities of components, osmotic coefficients, molar volumes and other thermodynamic parameters for the whole range of possible salinities (from zero up to approximately 10 molal at 573.15 K). This is a new result for NaCl water solutions. The range of the $TP$ parameters, covered by our model relates, in particular, to ore-forming H$_2$O-NaCl hydrothermal systems of epithermal, some part of porphyry, and the most part of orogenic gold deposits.

The further development of the approach presented in this paper should be directed to building similar thermodynamic models of other saline solutions than H$_2$O-NaCl. The fact, that our approach works well both for dilute and very concentrated solutions makes it prospective for highly solvable salts like CaCl$_2$. Calcium chloride is able to build much more concentrated solutions than NaCl, and these concentrated solutions play an important role in many geological processes (Bischoff *et al*, 1996). Development of such a model for $H_2O\text{-}CaCl_2$ is a current research of the authors.

An evident way of application of our model for $H_2O\text{-}NaCl$ is its incorporation into models of ternary systems $H_2O\text{-}NaCl$-non-polar gas, first of all the ternary system $H_2O\text{-}NaCl\text{-}CO_2$, which plays a key role in studying fluids in geological systems. In particular, this concerns the problem of formation of hydrothermal deposits. Our results presented above, can be used as an element for building the more precise models for these systems covering a broad range of salt concentrations, including solutions near to the saturation. Building such models is planned by the authors. The second way of application of results presented in the current paper is their direct usage in studying of the fluid mass transfer. It is known that the transportation ability of the fluid strongly depends on its density. *PTx* dependencies of density of $H_2O\text{-}NaCl$ fluids were studied in details in the works by Driesner and Heinrich (2007) and Driesner (2007). Our results on the densities of $H_2O\text{-}NaCl$ fluids are very near to those by Driesner (2007). On the other hand, the fluid-rock interaction depends on the activities of components of the fluid. Thus, our model for $H_2O\text{-}NaCl$ opens a possibility of enriching the analyses of the thermohaline convection in the Earth's crust (*e.g.* Geiger *et al*, 2006a, 2006b) with details concerning the fluid-rock interaction.

Our model for $H_2O\text{-}NaCl$ can be obtained in the form of a computer code by contacting the authors.

## 8. Acknowledgments

**Appendix**

The chemical potential of water, following from the equation for the excess Gibbs free energy (13) is the sum of corresponding partial chemical potentials, which are

$$\mu_1^{id} = RT \ln x_1 \tag{A1}$$

$$\mu_{1,\alpha} = -RT \ln(1 + \alpha x_2) \tag{A2}$$

$$\mu_{1,a} = \tfrac{1}{2}\left\{x_2^{1/2} \ln\left(1 + x_2^{1/2}/\varepsilon_a\right) - x_2/\left(\varepsilon_a + x_2^{1/2}\right)\right\} W_a \tag{A3}$$

$$\mu_{1,2} = x_2(1 - x_1) W_2 \tag{A4}$$

$$\mu_{1,6} = x_2^2(1 - 2x_1) W_6 \tag{A5}$$

Corresponding partial chemical potentials of NaCl, necessary for calculation of $\gamma_\pm$ are the following:

$$\mu_2^{id} = RT \ln x_2 \tag{A6}$$

$$\mu_{2,\alpha} = RT\{\alpha \ln x_2 + (1+\alpha)\ln(1+\alpha) - (1+\alpha)\ln(1+\alpha x_2)\} \tag{A7}$$

$$\mu_{2,a} = \left\{\frac{1+x_2}{2x_2^{1/2}} \ln\left(1 + x_2^{1/2}/\varepsilon_a\right) + \frac{1-x_2}{2(\varepsilon_a + x_2^{1/2})} - \ln(1 + 1/\varepsilon_a)\right\} W_a \tag{A8}$$

$$\mu_{2,2} = x_1(1 - x_2) W_2 \tag{A9}$$

$$\mu_{2,6} = 2 x_1 x_2 (1 - 2x_2) W_6 \tag{A10}$$

The terms contributing into $\mu_{0,2}$ are:

$$\mu_{0,2,\alpha} = RT(1+\alpha)\ln(1+\alpha) \tag{A11}$$

$$\mu_{0,2,a} = \{1/\varepsilon_a - \ln(1 + 1/\varepsilon_a)\} W_a \tag{A12}$$

$$\mu_{0,2,2} = W_2 \tag{A13}$$

In the equations given above, we did not use equality $x_1 + x_2 = 1$ to retain these formulas more convenient for a transformation for systems with more than two components.